\documentclass{article}

\usepackage{microtype}
\usepackage{graphicx}
\usepackage{subfigure}
\usepackage{booktabs} 
\usepackage[table]{xcolor}
\usepackage{graphicx}
\usepackage{textcomp}

\usepackage{hyperref}

\usepackage[accepted]{icml2020}

\icmltitlerunning{COVID-19 Image Data Collection}

\begin{document}

\twocolumn[
\icmltitle{COVID-19 Image Data Collection}



\icmlsetsymbol{equal}{*}

\begin{icmlauthorlist}
\icmlauthor{Joseph Paul Cohen}{mila,udem}
\icmlauthor{Paul Morrison}{f}
\icmlauthor{Lan Dao}{fm}
\end{icmlauthorlist}

\icmlaffiliation{mila}{Mila, Quebec Artificial Intelligence Institute}
\icmlaffiliation{udem}{University of Montreal}
\icmlaffiliation{fm}{Faculty of Medicine, University of Montreal}
\icmlaffiliation{f}{Department of Mathematics and Computer Science, Fontbonne University}

\icmlcorrespondingauthor{Joseph Paul Cohen}{joseph@josephpcohen.com}

\icmlkeywords{Computer Vision, COVID-19}

\vskip 0.3in
]



\printAffiliationsAndNotice{}  

\begin{abstract}
This paper describes the initial COVID-19 open image data collection. It was created by assembling medical images from websites and publications and currently contains 123 frontal view X-rays.
\end{abstract}

\section{Motivation}

In the context of a COVID-19 pandemic, is it crucial to streamline diagnosis. Data is the first step to developing any diagnostic tool or treatment. While there exist large public datasets of more typical chest X-rays \cite{WangNIH2017, Bustos2019,Irvin2019,Johnson2019, Demner-Fushman2016}, there is no collection of COVID-19 chest X-rays or CT scans designed to be used for computational analysis.

\begin{figure}[!th]
\centering
    \subfigure[Day 10]{
    \includegraphics[width=0.46\columnwidth]{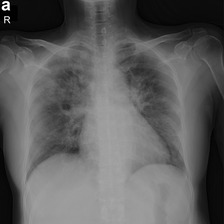}
    }%
    \subfigure[Day 13]{
    \includegraphics[width=0.46\columnwidth]{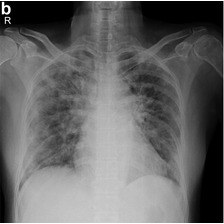}
    }%
    
    \subfigure[Day 17]{
    \includegraphics[width=0.46\columnwidth]{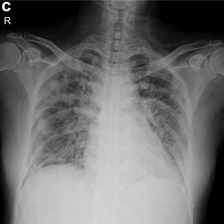}
    }%
    \subfigure[Day 25]{
    \includegraphics[width=0.46\columnwidth]{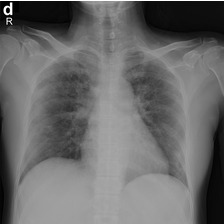}
    }%
    
    \caption{Example images from the same patient (\#19) extracted from \citet{10.1016/j.jfma.2020.02.007}. This 55 year old female survived a COVID-19 infection.}
    \label{fig:example}
\end{figure}

In this paper, we describe the public database of pneumonia cases with chest X-ray or CT images, specifically COVID-19 cases as well as MERS, SARS, and ARDS. Data will be collected from public sources in order not to infringe patient confidentiality. Example images shown in Figure \ref{fig:example}.

Our team believes that this database can dramatically improve identification of COVID-19. Notably, this would provide essential data to train and test a Deep Learning-based system, likely using some form of transfer learning. These tools could be developed to identify COVID-19 characteristics as compared to other types of pneumonia or in order to predict survival.

Currently, all images and data are released under the following URL: \url{https://github.com/ieee8023/covid-chestxray-dataset}. As stated above, images collected have already been made public.

\begin{figure*}[!th]
\centering
\includegraphics[width=0.33\textwidth]{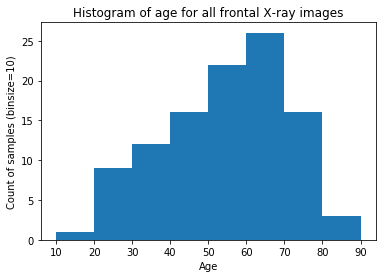}%
\includegraphics[width=0.33\textwidth]{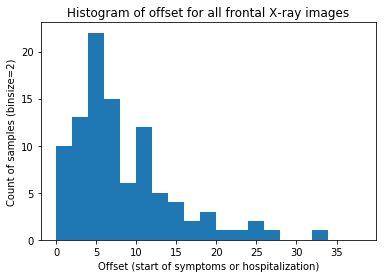}%
\includegraphics[width=0.33\textwidth]{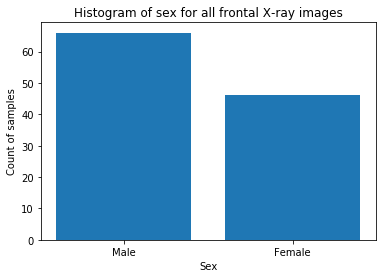}

    \caption{Demographics for each frontal X-ray image}
    \label{fig:demographics}
\end{figure*}

\begin{table}[]
    \centering
    \begin{tabular}{c c c c}
\toprule
Type of pneumonia      & PA & AP & AP Supine\\
\midrule
SARSr-CoV-2 or COVID-19     & 76 & 11 & 13\\
SARSr-CoV-1 or SARS         & 11 & 0  & 0\\
\textit{Streptococcus spp.} & 6  & 0  & 0\\
\textit{Pneumocystis spp.}  & 1  & 0  & 0\\
ARDS          & 4  & 0  & 0\\
\bottomrule
    \end{tabular}
    \caption{Counts of each finding and view. PA = posteroanterior, AP = anteroposterior, AP Supine = laying down, SARSr-CoV-2 = Severe acute respiratory syndrome-related coronavirus 2, SARS-CoV-1 or SARS = Severe acute respiratory syndrome-related coronavirus 1, ARDS = acute respiratory distress syndrome}
    \label{tab:stats}
\end{table}

\begin{table}[!th]
\rowcolors{2}{gray!5}{white}
    \centering
    \begin{tabular}{c p{151pt}}
\toprule
Attribute      & Description \\
\midrule
Patient ID& Internal identifier\\

Offset& Number of days since the start of symptoms or hospitalization for each image. If a report indicates "after a few days", then 5 days is assumed.\\

Sex& Male (M), Female (F), or blank\\

Age& Age of the patient in years\\

Finding &Type of pneumonia\\

Survival &Yes (Y) or no (N)\\

View&Posteroanterior (PA), Anteroposterior (AP), AP Supine (APS), or Lateral (L) for X-rays; Axial or Coronal for CT scans\\

Modality& CT, X-ray, or something else\\

Date& Date on which the image was acquired\\

Location& Hospital name, city, state, country\\

Filename& Name with extension\\

doi & Digital object identifier (DOI) of the research article\\

url & URL of the paper or website where the image came from\\

License& License of the image such as CC BY-NC-SA. Blank if unknown \\

Clinical notes& Clinical notes about the image and/or the patient \\

Other notes &e.g. credit \\

\bottomrule
    \end{tabular}
    \caption{Descriptions of each attribute of the metadata}
    \label{tab:desc}
\end{table}

\section{Expected outcome}

This dataset can be used to study the progress of COVID-19 and how its radiological findings vary from other types of pneumonia. Similarly to the outcome of the Chest Xray14 \cite{WangNIH2017} dataset which enabled significant advances in medical imaging, tools can be developed to predict not only the type of pneumonia, but also its outcome. Eventually, our model could take inspiration from work by \citet{Rajpurkar2017a}, which could predict pneumonia, as well as \citet{Cohen2019}, which deployed such models.

Tools could be built to triage cases in the absence of physical tests, particularly in the context of polymerase chain reaction (PCR) tests shortage \cite{Satyanarayana2020, KellyGeraldineMalone2020}. These tools could predict patient outcomes such as survival, allowing a physician to plan ahead for specific patients and facilitate management. In extreme situations, where physicians could be faced with the extraordinary decision to choose which patient should be allocated healthcare resources \cite{YaschaMounk2020}, such a tool could potentially serve as a measuring device. 

Furthermore, these tools could monitor the progression of COVID-19 positive patients in order to better track the evolution of their condition. Ultimately, this dataset and its analysis could help us better understand the dynamics of the disease and better prepare treatments.

\vspace{-5pt}
\section{Dataset}
\vspace{-5pt}
The current statistics as of March 25th 2020 are shown in Table \ref{tab:stats}.
For each image, attributes shown in Table \ref{tab:desc} are collected.
Data is largely compiled from websites such as Radiopaedia.org, the Italian Society of Medical and Interventional Radiology\footnote{https://www.sirm.org/category/senza-categoria/covid-19/}, and Figure1.com\footnote{https://www.figure1.com/covid-19-clinical-cases}. Images are extracted from online publications, website, or directly from the PDF using the tool pdfimages\footnote{https://poppler.freedesktop.org/}. The goal during this process is to maintain the quality of the images.

Data was collected from the following papers:  \cite{10.1056/nejmc2001272,
10.1056/NEJMc2001573,
10.1016/S0140-67362030211-7,
10.1148/rg.242035193,
10.1016/S0140-67362030370-6,
10.1148/radiol.2020200269,
10.1056/NEJMoa2001191,
10.1148/ryct.2020200034,
10.1148/ryct.2020200028,
10.3346/jkms.2020.35.e79,
10.1148/radiol.2020200490,
10.1016/j.jfma.2020.02.007,
10.1111/all.14238,
10.1016/j.jmii.2020.03.003,
10.1093/cid/ciaa199,
10.3348/kjr.2020.0132,
10.1016/j.jmii.2020.03.008,
10.1016/S1473-30992030111-0,
10.1038/s41591-020-0819-2,
10.3348/kjr.2020.0112}

\bibliography{refs,chester-covid-19,papers}
\bibliographystyle{icml-nopagenum}

\end{document}